\documentclass[manuscript]{emulateapj}

\shorttitle{Confined solar flares observed by NVST and SDO}
\shortauthors{Yang et al.}

\journalinfo{Accepted for publication in ApJL}

\submitted{Accepted for publication in ApJL}

\begin{document}

\title{Fine Structures and Overlying Loops of Confined Solar Flares}

\author{Shuhong Yang\altaffilmark{1}, Jun Zhang\altaffilmark{1}, and
Yongyuan Xiang\altaffilmark{2}}

\altaffiltext{1}{Key Laboratory of Solar Activity, National
Astronomical Observatories, Chinese Academy of Sciences, Beijing
100012, China; shuhongyang@nao.cas.cn}

\altaffiltext{2}{Fuxian Solar Observatory, Yunnan Observatories,
Chinese Academy of Sciences, Kunming 650011, China}

\begin{abstract}

Using the H$\alpha$ observations from the New Vacuum Solar Telescope
at \emph{Fuxian Solar Observatory}, we focus on the fine structures
of three confined flares and the issue why all the three flares are
confined instead of eruptive. All the three confined flares take
place successively at the same location and have similar
morphologies, so can be termed homologous confined flares. In the
simultaneous images obtained by the \emph{Solar Dynamics
Observatory}, many large-scale coronal loops above the confined
flares are clearly observed in multi-wavelengths. At the pre-flare
stage, two dipoles emerge near the negative sunspot, and the dipolar
patches are connected by small loops appearing as arch-shaped
H$\alpha$ fibrils. There exists a reconnection between the small
loops, and thus the H$\alpha$ fibrils change their configuration.
The reconnection also occurs between a set of emerging H$\alpha$
fibrils and a set of pre-existing large loops, which are rooted in
the negative sunspot, a nearby positive patch, and some remote
positive faculae, forming a typical three-legged structure. During
the flare processes, the overlying loops, some of which are tracked
by activated dark materials, do not break out. These direct
observations may illustrate the physical mechanism of confined
flares, i.e., magnetic reconnection between the emerging loops and
the pre-existing loops triggers flares and the overlying loops
prevent the flares from being eruptive.

\end{abstract}


\section{INTRODUCTION}

Solar flares are one of the most energetic phenomena in the solar
atmosphere, indicating the explosive release of a great deal of
energy, and one mechanism of the flare energy release is believed to
be magnetic reconnection by many solar researchers. Flares and
coronal mass ejections (CMEs) are suggested to be two different
manifestations of the energy release process, as revealed by the
previous observations (Harrison 1995; Zhang et al. 2001a). However,
not all the flares are accompanied by CMEs, which can eject
large-scale or even global magnetic structures into the
interplanetary space (Harrison 1995; Yashiro et al. 2005). The
flares associated with CMEs are termed ``eruptive flares" and the
others ``confined flares" (Svestka \& Cliver 1992). The eruptive
flares usually last for a long period, from tens of minutes to
hours, while the confined ones have a short duration. The occurrence
rate of eruptive flares is dependent on the flare intensity and
duration (Kahler et al. 1989; Andrews 2003). Generally, the larger
flares in energy with longer duration tend to be eruptive. Wang \&
Zhang (2007) found that the occurrence of eruption (or confinement)
is sensitive to the distance between the flares and the active
regions, 22-37 Mm for eruptive flares and 6-17 Mm for confined
flares. On the other hand, not all the CMEs are associated with
noticeable flares (St.~Cyr \& Webb 1991; Green et al. 2002; Wang et
al. 2002; Zhou et al. 2003; Ma et al. 2010).

To explain the physical mechanism of eruptive events, many theories
and models have been proposed, in most of which there should exist
an opening process of the overlying magnetic loops so that plasma
and magnetic flux can escape (Lin et al. 2003; Forbes et al. 2006).
For the eruptive flares, a standard model which involves a rising
flux rope stretching the overlying magnetic lines has been widely
accepted (Sturrock 1966; Masuda et al. 1994; Shibata et al. 1995;
Tsuneta 1996). According to this model, between the stretched
antiparallel lines, a current sheet is formed and magnetic
reconnection takes place. Thus the released energy heats the coronal
plasma and also accelerates particles. About the onset of eruptions,
many different initiation mechanisms, such as flux emergence and
cancellation, breakout, tether cutting, torus instability, and kink
instability, have been proposed and well studied (Forbes \& Isenberg
1991; Antiochos et al. 1999; Chen \& Shibata 2000; Moore et al.
2001; Zhang et al. 2001b; T{\"o}r{\"o}k \& Kliem 2005, 2007). The
confined flares are mainly affected by the surrounding coronal
magnetic fields. Simulations by T{\"o}r{\"o}k \& Kliem (2005) and
Fan \& Gibson (2007) revealed that the slow decrease of the
overlying arcade field with height is a major factor in permitting
the instability to result in a confined event. Based on the
potential field source-surface model, the calculations showed that
stronger overlying magnetic arcades can prevent energy release, thus
resulting in confined flares (Wang \& Zhang 2007; Guo et al. 2010;
Cheng et al. 2011). Direct observations of the overlying arcades
above the failed filament eruptions have been provided by many
authors (e.g., Ji et al. 2003; Zheng et al. 2012; Chen et al. 2013).

In the previous studies, direct observations for the overlying loops
above confined flares and the reconnection process between H$\alpha$
fibrils are rare. In this Letter, we mainly study the fine
structures and the overlying loops of three homologous confined
flares simultaneously observed by both the New Vacuum Solar
Telescope (NVST; Liu et al. 2014) and the \emph{Solar Dynamics
Observatory} (\emph{SDO}; Pesnell et al. 2012). The NVST is a
primary facility of the \emph{Fuxian Solar Observatory}, which is
located on the northeast side of Fuxian Lake in China. Since the
solar chromosphere is highly dynamic (van Noort \& Rouppe van der
Voort 2006; Lin et al. 2009; Yang et al. 2014), the high
tempo-spatial resolution H$\alpha$ observations with NVST can
provide plentiful information of dynamic events, e.g., rapid changes
of H$\alpha$ fibrils during solar flares. The \emph{SDO} provides
full-disk multi-wavelength images of the Sun and high-quality
magnetograms of the photosphere. The combination of these
observations can help us to study the physical properties of
confined flares in detail.

\begin{figure*}
\centering
\includegraphics
[bb=43 169 543 665,clip,angle=0,width=0.95\textwidth]
{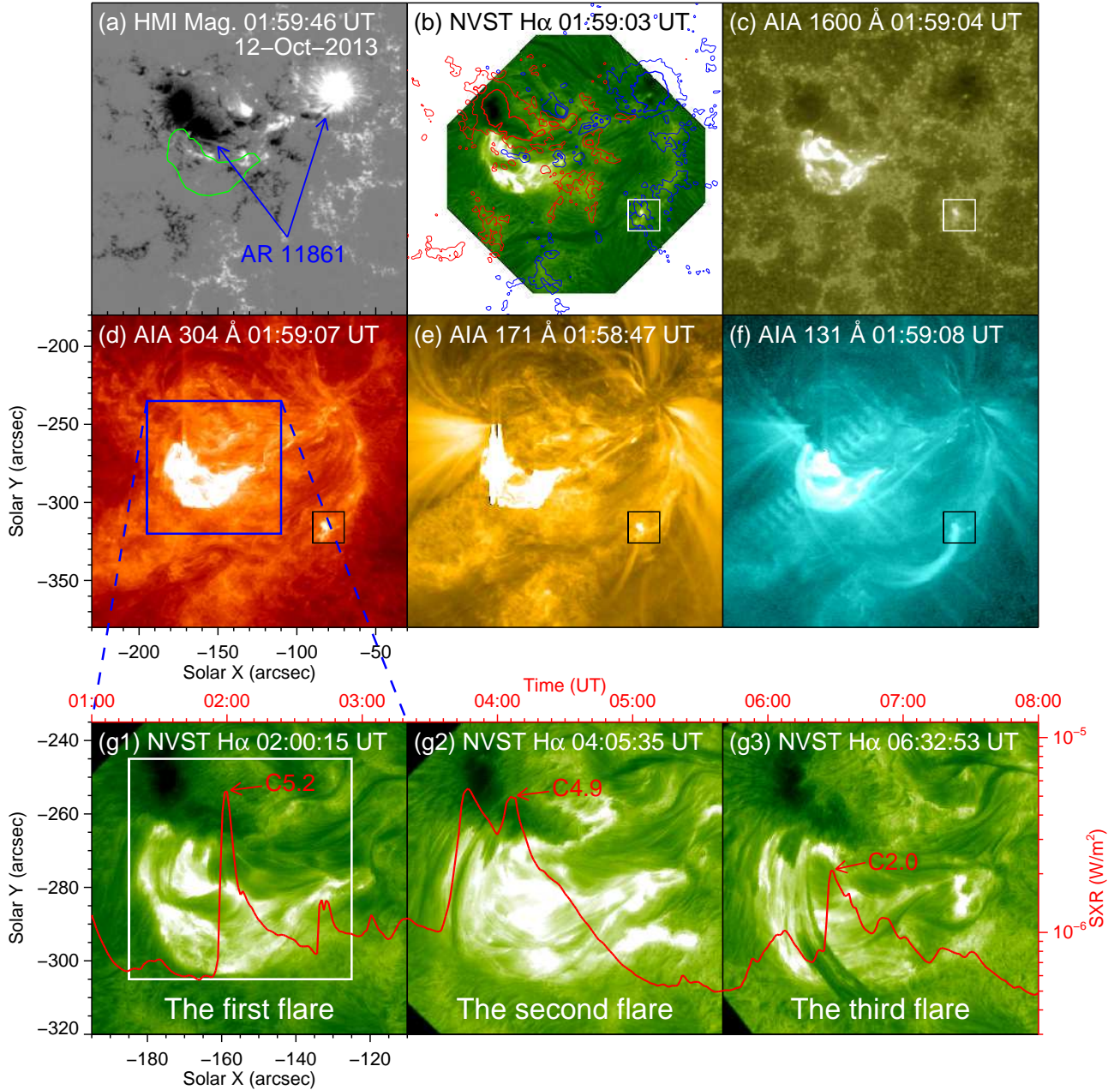} \caption{Panels (a)-(f): HMI line-of-sight
magnetogram, NVST H$\alpha$, AIA 1600 {\AA}, 304 {\AA}, 171 {\AA},
and 131 {\AA} images displaying the overview of AR 11861 where the
homologous confined flares occurred. Panels (g1)-(g3): expanded
H$\alpha$ images showing the three confined flares taking place
successively, and the overplotted red curve shows the variation of
the \emph{GOES} soft X-ray flux. The green curve in panel (a)
outlines the general shape of the first flare, and the red and blue
curves in panel (b) are the contours of the corresponding positive
and negative magnetic fields at levels of (20, 80) G and (-20, -80)
G, respectively. The small boxes in panels (b)-(f) outline the
brightening region, and the white square in panel (g1) outlines the
FOV of Figure 2. \label{fig}}
\end{figure*}

\begin{figure*}
\centering
\includegraphics
[bb=80 217 479 604,clip,angle=0,width=0.9\textwidth] {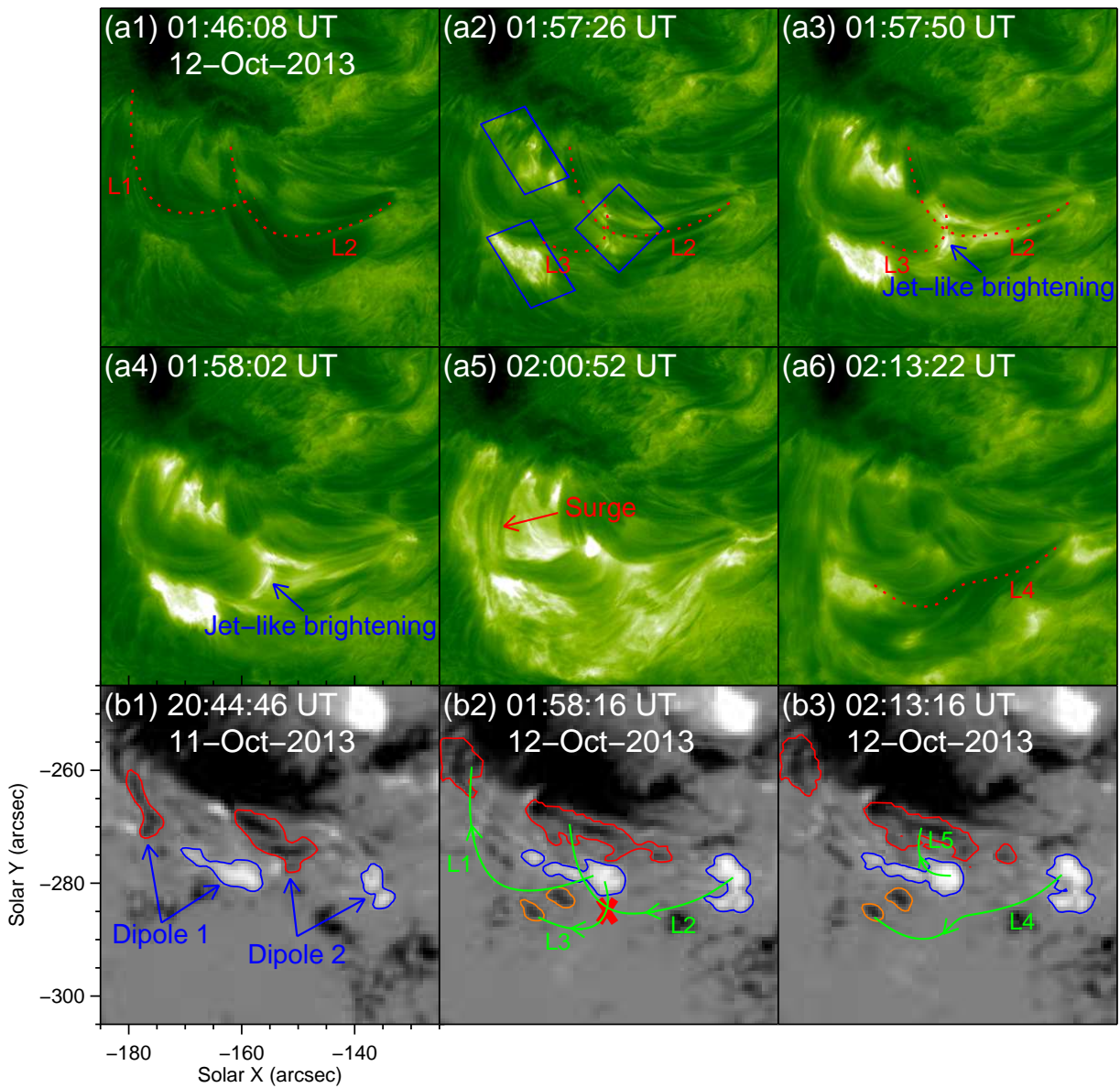}
\caption{ Panels (a1)-(a6): NVST H$\alpha$ images displaying the
process of the first flare (also see Movie 1). Panels (b1)-(b3): HMI
magnetograms showing the evolution of the underlying magnetic fields
(also see Movie 2). Curves ``L1"--``L4" represent the visible loops
identified in the H$\alpha$ images, while curve ``L5" in panel (b3)
represents the invisible low-lying loops newly formed due to the
reconnection. The red cross symbol in panel (b2) marks the site
where the reconnection occurred. The quadrilaterals in panel (a2)
outline the areas with initial brightenings. \label{fig}}
\end{figure*}

\begin{figure*}
\centering
\includegraphics
[bb=100 242 466 585,clip,angle=0,width=0.8\textwidth] {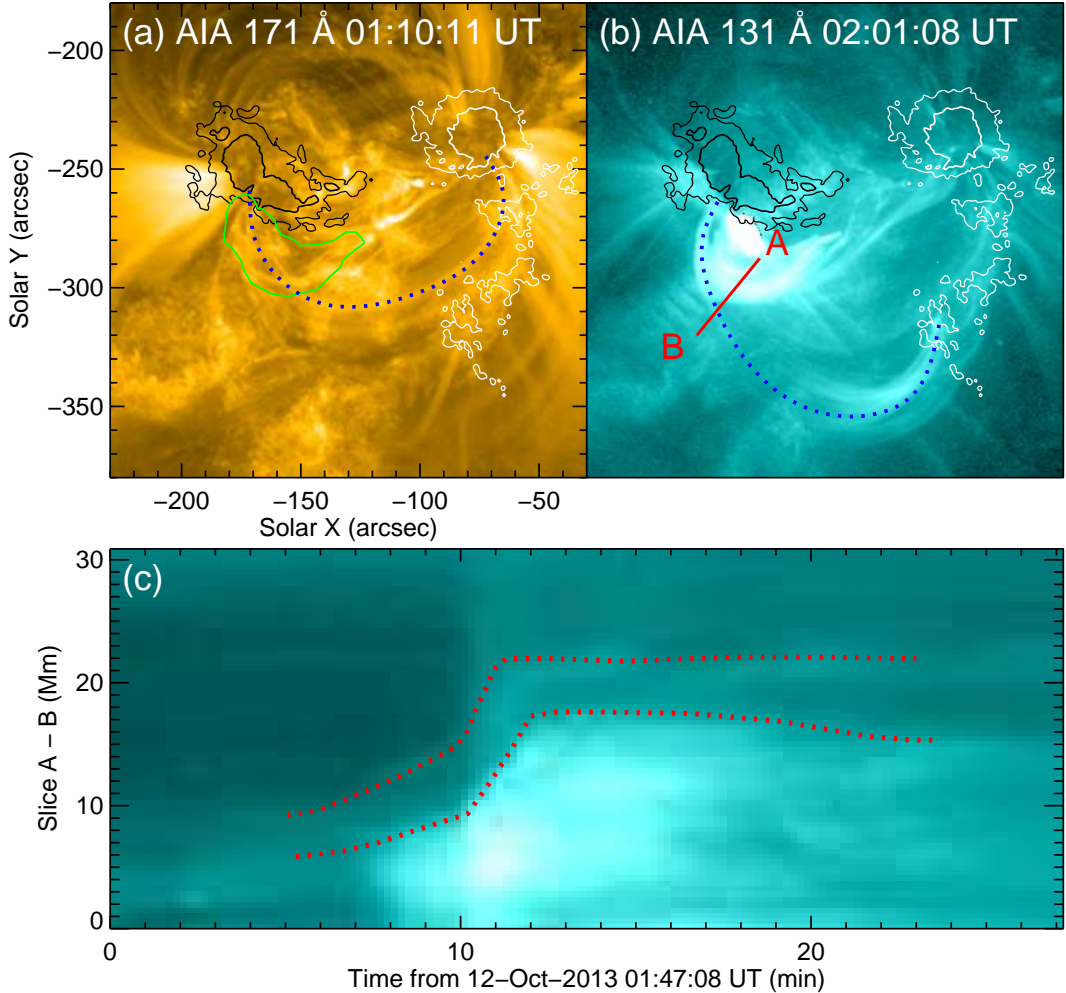}
\caption{Panels (a)-(b): AIA 171 {\AA} and 131 {\AA} images showing
the overlying coronal loops above the first confined flare (also see
Movie 3). Panel (c): space-time plot along slice ``A--B" marked in
panel (b). The green curve in panel (a) outlines the shape of the
first flare. The dotted curves in panels (a) and (b) delineate the
most conspicuous loops, and the white and black curves are the
contours of the positive and negative fields where the loops are
anchored. The dotted curves in panel (c) indicate the evolution of
the overlying loop and the edge of the flare observed in 131 {\AA}.
\label{fig}}
\end{figure*}

\begin{figure*}
\centering
\includegraphics
[bb=70 150 489 694,clip,angle=0,width=0.75\textwidth]
{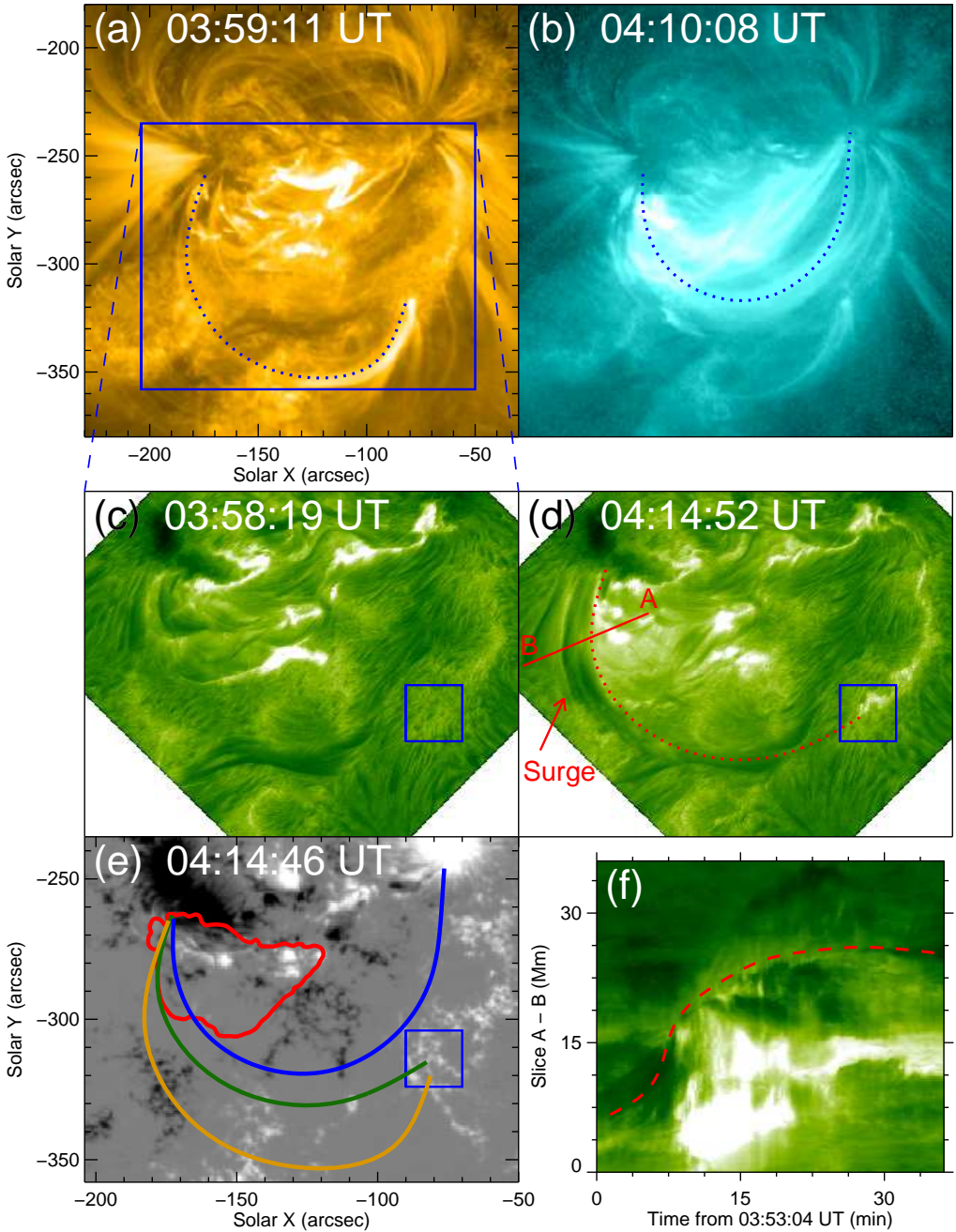} \caption{Panels (a)-(b): AIA 171 {\AA} and 131
{\AA} images showing the coronal loops above the second confined
flare. Panels (c)-(e): NVST H$\alpha$ images and HMI magnetogram
showing the confined flare (also see Movie 4) and the underlying
magnetic fields. Panel (f): space-time plot along slice ``A--B"
marked in panel (d). The curves in panels (a), (b), and (d)
delineate the loops identified in 171 {\AA}, 131 {\AA}, and
H$\alpha$ images, which are also overlaid in panel (e) with brown,
blue, and green curves. The squares in panels (c)-(e) outline the
area with chromospheric brightenings. The red curve in panel (e)
outlines the general shape of the second flare. The dashed curve in
panel (f) indicates the overlying arcade above the second flare.
\label{fig}}
\end{figure*}

\begin{figure*}
\centering {\includegraphics [bb=47 226 513
593,clip,angle=0,width=0.75\textwidth] {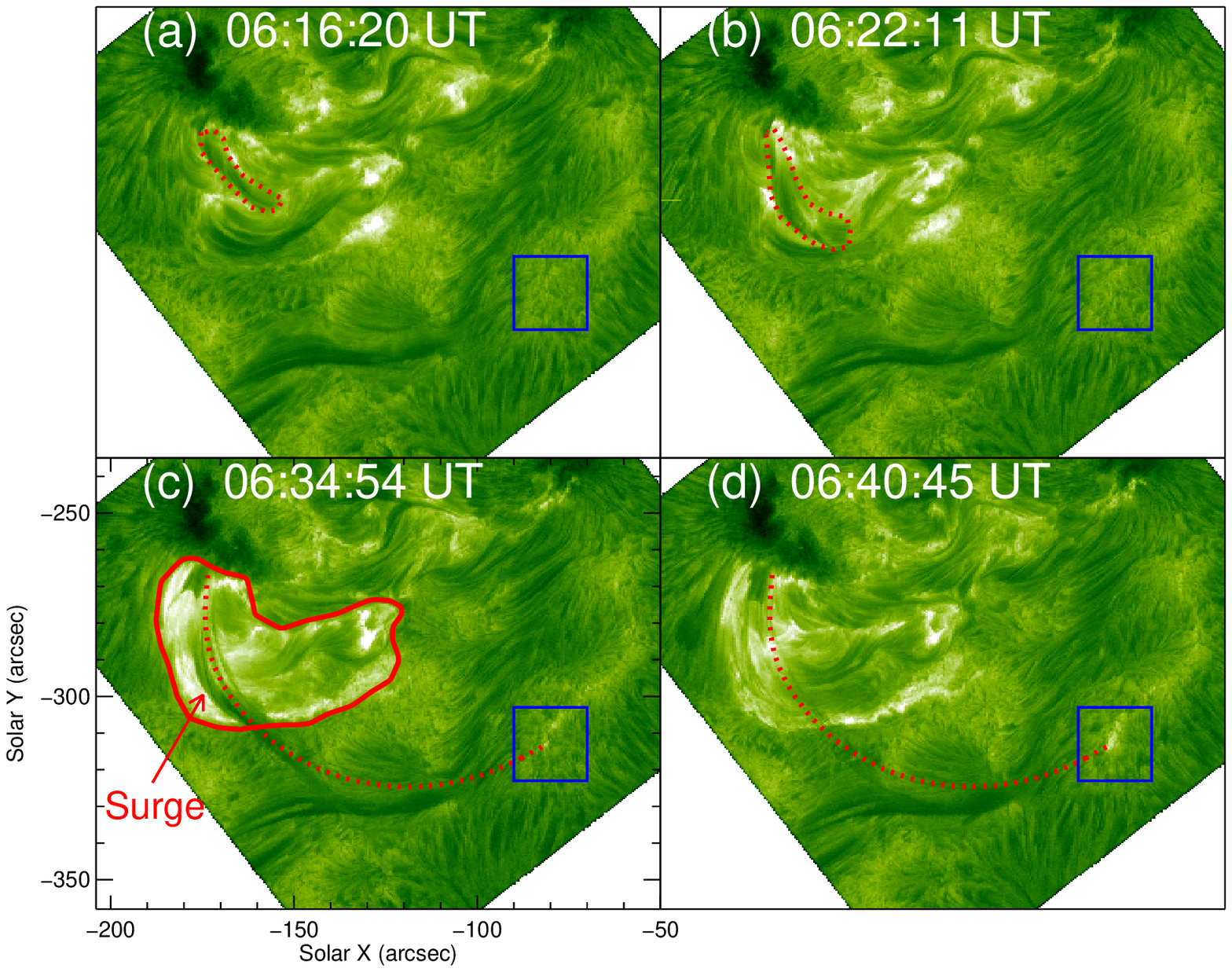}}
{\includegraphics [bb=47 330 513
508,clip,angle=0,width=0.75\textwidth] {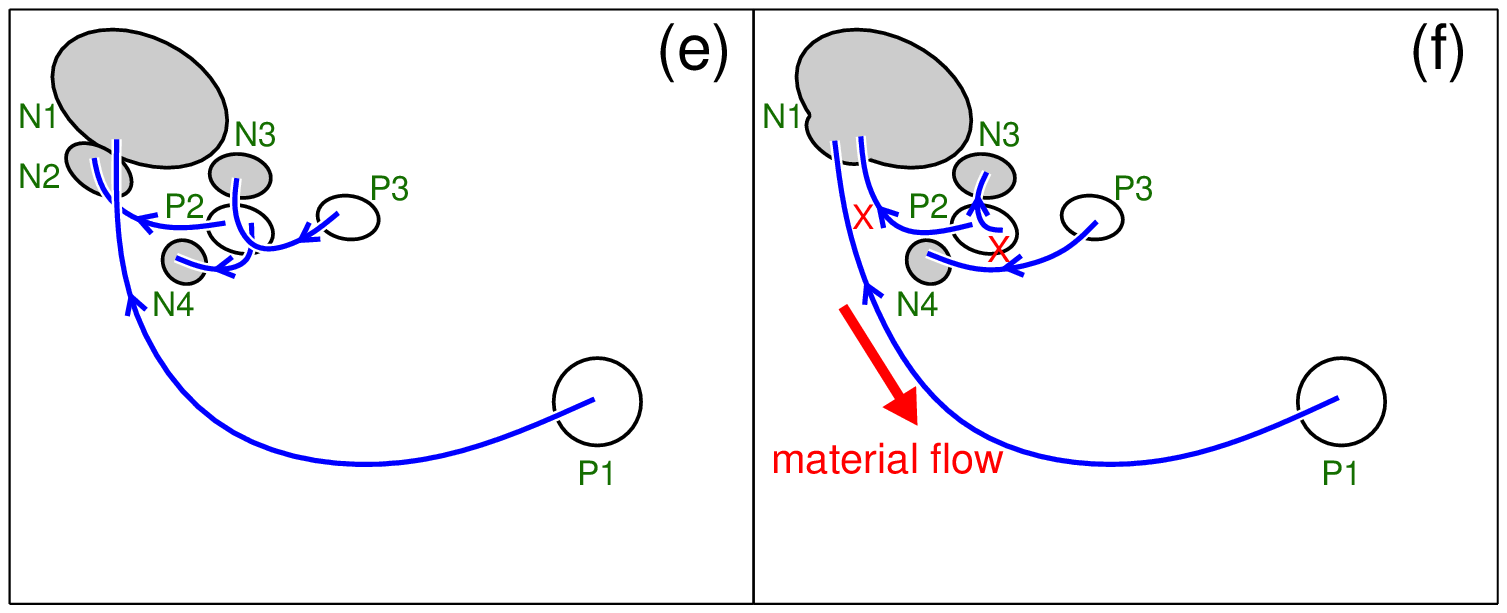}} \caption{Panels
(a)-(d): NVST H$\alpha$ images showing the third confined flare
(also see Movie 5). The red curves and blue squares are similar to
those in Figure 4. Panels (e)-(f): schematic drawings illustrating
the magnetic configurations of the confined flares before and after
magnetic reconnection. The filled and open ellipses represent the
negative and positive magnetic patches, respectively. The blue
curves with arrows show the loop connections between the opposite
polarities, and the red cross symbols indicate the reconnection
sites. \label{fig}}
\end{figure*}

\section{OBSERVATIONS AND DATA ANALYSIS}

The NVST has three channels being used to image the Sun, i.e.,
H$\alpha$, TiO, and G band. The channel for observing the solar
chromosphere is H$\alpha$ 6562.8 {\AA} with a bandwidth of 0.25
{\AA}. It can be tuned in the range of $\pm$5 {\AA} with a step of
0.1 {\AA}. The NVST data adopted here were obtained in H$\alpha$
6562.8 {\AA} from 01:01:32 UT to 09:10:42 UT on 2013 October 12 with
a cadence of 12 s. In the sequence of H$\alpha$ images, there is a
gap from 05:40:26 UT to 05:49:54 UT. The H$\alpha$ observations
cover most of the AR 11861 with a field-of-view (FOV) of 152$''$
$\times$ 152$''$ and a pixel size of 0$\arcsec$.164. The data are
first calibrated, including dark current subtraction and flat field
correction, and then reconstructed by speckle masking (Weigelt 1977;
Lohmann et al. 1983). In order to understand the flares better, the
\emph{Geostationary Operational Environmental Satellite}
(\emph{GOES}) data are employed to show the variation of soft X-ray
flux.

Moreover, we also use the Helioseismic and Magnetic Imager (HMI;
Scherrer et al. 2012; Schou et al. 2012) magnetograms and the
Atmospheric Imaging Assembly (AIA; Lemen et al. 2012)
multi-wavelength images from the \emph{SDO}. AIA 1600 {\AA}, 304
{\AA}, 171 {\AA}, and 131 {\AA} data from 01:00 UT to 09:20 UT on
October 12 with a spatial sampling of 0$\arcsec$.6 pixel$^{-1}$ are
chosen. The cadence of the 1600 {\AA} images is 24 s, and that of
the other three wavelengths is 12 s. The HMI line-of-sight
magnetograms were observed from 12:00 UT on October 11 to 12:00 UT
on October 13. They have a pixel size of 0$\arcsec$.5 and a cadence
of 45 s. The AIA and HMI data are co-aligned by applying the
standard routine \emph{aia\_prep.pro} available in the Solar
Software (SSW) package, and differentially rotated to a reference
time (02:00 UT, 2013 October 12). The \emph{SDO} and NVST data are
aligned using the cross-correlation with specific features.

\section{RESULTS}

On 2013 October 12, the NVST observed three confined flares in good
seeing conditions. The flares occurred successively in NOAA AR
11861, which is a complex sunspot group with a type of
$\beta\gamma\delta$ (see Figure 1(a)). All the three flares were
located at the same location, close to the main sunspot with
negative polarity. Each of the flares can be observed in
multi-wavelengths. Panels (b)-(f) show the overview of the first
flare observed in five wavelengths, 6562.8 {\AA}, 1600 {\AA}, 304
{\AA}, 171 {\AA}, and 131 {\AA}, which are formed in the different
layers of the solar atmosphere, from the chromosphere to the corona.
We can see that the first flare appears as a compact bright
structure. The general appearances of the three well-developed
confined flares are shown in panels (g1)-(g3), respectively. These
flares can be termed homologous confined flares since they (1) occur
at an identical location, (2) have similar morphologies, (3) and
take place successively with time intervals of 2.1 hr and 2.37 hr,
and (4) their classes (C5.2, C4.9, and C2.0) are comparable (see the
overplotted red curve in the bottom panels).

For the first flare, its evolution process is shown in Figure 2
(also see Movie 1). At the pre-flare stage, many small loops
outlined by arch-shaped H$\alpha$ fibrils can be clearly observed,
as delineated with curves ``L1'' and ``L2'' (panel (a1)). Then in
some areas near the footpoints and the interface of loops, several
significant chromospheric brightenings appeared (outlined by
quadrilaterals in panel (a2)), indicating the onset of the flare. At
that time, some other loops (outlined by curve ``L3") were detected.
At the intersection of ``L2" and ``L3", a jet-like brightening
(denoted by arrows in panels (a3) and (a4)) was observed, which can
be considered as a signature of reconnection. The flare went on
develop and reached the maximum around 02:00 UT (see panel (a5)).
About 13 min later, the flare decayed significantly and a new loop
configuration (labeled with ``L4") was formed (panel (a6)). During
this process, there exists an ejection of dark material, appearing
as a dark surge (denoted by the arrow in panel (a5)) which can be
well identified from 01:58 UT to 02:15 UT. The bottom panels in
Figure 2 display the evolution of the underlying magnetic fields
(also see Movie 2). Before the occurrence of the first flare, two
dipoles (indicated by arrows in panel (b1)) emerged and separated
near the negative sunspot. There exists a great shear motion between
the two dipoles during their emergence. For both of the dipoles, the
magnetic patches with negative polarity moved toward the negative
sunspot, while the positive ones away from it. In panels (b2)-(b3),
the curves ``L1"-``L4" overlaid in the magnetograms are the loops
identified in the corresponding H$\alpha$ images. As shown in panel
(b2), loops ``L1" and ``L2" connect the opposite polarities of
dipoles ``1" and ``2", respectively. There also exists a loop
connection ``L3" between the positive polarity of dipole ``1" and
the nearby negative fields. Then magnetic reconnection took place
between loops ``L2" and ``L3" at the site marked by the red cross
symbol. Due to the reconnection, loops ``L4" and ``L5" were newly
formed, as indicated by the green curves in panel (b3). It seems
that the newly formed loop ``L4" (``L5") should be from the positive
polarity of dipole ``2" (dipole ``1") to the negative polarity of
the other dipole. Loop ``L4" is quite obvious in the H$\alpha$ image
shown in panel (a6), while the low-lying loop ``L5" is speculated to
exist although it cannot be identified in the corresponding
H$\alpha$ image.

In the 171 {\AA} and 131 {\AA} images observed by AIA, many long and
high coronal loops are observed above the flare (see Figures
3(a)-(b) and Movie 3). 171 {\AA} line corresponds to a low
temperature (0.6 MK), while 131 {\AA} line corresponds to a high
temperature (11 MK). Here, 171 {\AA} and 131 {\AA} images are used
to show the appearance of the overlying loops at different
temperatures. The dotted curves in panels (a) and (b) delineate the
most conspicuous loops in 171 {\AA} and 131 {\AA} images,
respectively. They are large-scale loops with an average length of
about 130 Mm. Most of them connect the two main sunspots, the
leading one with positive polarity and the following one with
negative polarity. Besides, there are also many loops connecting the
negative sunspot and the faculae with the positive polarity. Along
slice ``A--B" marked in panel (b), we make a space-time plot of a
sequence of 131 {\AA} images to study the evolution of the first
flare and overlying loops and present it in panel (c). The higher
and lower dotted curves indicate the leading edges of the loop and
the first flare, respectively. As revealed by the two curves, the
coronal loops lifted at first as the flare expanded, and then
stopped at a certain height when the flare began to decay. Both of
them have a similar evolution process at first, a slow-rise phase
and a consequent fast-rise phase, after which the coronal loop
stayed at the final location while the flare began to decrease.

About 2 hr later after the first flare onset, the second confined
flare occurred (see Movie 4). Similar to the first one, the second
flare took place at the same site and many overlying coronal loops
existed at the pre-flare stage (Figures 4(a)-(b)). The coronal loops
maintained their initial configuration without significant change.
In H$\alpha$ images, a lot of dark materials were activated and
ejected, appearing as a dark surge, and thus tracked a large
arch-shaped structure which is emphasized with the red curve (panels
(c)-(d)). One footpoint of the tracked structure is located within
the negative sunspot, and the other one is at the quiet region about
60 Mm away (outlined by the small squares). At the pre-flare stage,
there was no obvious activity in the square region. While when the
activated dark materials reached the footpoint, conspicuous
chromospheric brightenings were observed. The overlying arcades
above the second flare identified in 171 {\AA}, 131 {\AA}, and
H$\alpha$ are plotted on the photospheric magnetogram with brown,
blue, and green curves, respectively (see panel (e)). We can see
that these loops are above the confined flare (indicated by the red
solid curve). We also obtain a space-time plot along slice ``A--B"
marked in panel (d) and present it in panel (f). As shown by the
dashed curve, the H$\alpha$ arcade rose at first and then stopped
gradually, instead of breaking out.

Figures 5(a)-(d) show the evolution of the third confined flare,
which took place about another 2 hr later after the second one (also
see Movie 5). Before the flare occurrence, a group of dark fibrils
began to rise and expand slowly (outlined by the red curves in
panels (a)-(b)). Meanwhile, the third flare took place at the same
site where the former two flares occurred, and all the three flares
generally have a similar shape. Then a dark surge appeared and some
of the dark materials moved along a large arcade (indicated by the
dotted curves in panels (c)-(d)), and ultimately fell down to the
solar surface (outlined by the squares) and the chromosphere
brightened up. The square region is almost the same location of the
brightenings observed during the first two flares (e.g., see the
small square area in Figure 1 and Movie 1).

\section{CONCLUSIONS AND DISCUSSION}

Based on the high resolution H$\alpha$ data obtained by the NVST, we
focus on the fine structures and the evolution processes of three
confined flares which successively occurred. The flares have similar
morphologies and the same location, and thus can be called
homologous confined flares. Many large-scale coronal loops above the
confined flares can be clearly identified in the AIA EUV images, and
some loops were tracked by dark materials observed in the NVST
H$\alpha$ images. Before the flares, two dipoles emerged near the
negative sunspot, and the opposite patches were connected by
arch-shaped fibrils. There exists a reconnection between two groups
of dark fibrils, resulting in the change of their configuration and
the energy release. The reconnection also took place between a set
of emerging small loops and a set of pre-existing large loops,
rooted in the negative sunspot, a nearby positive patch, and the
remote positive faculae. During the confined flares, the overlying
loops did not break out.

To illustrate the confined flares observed in this study, two
schematic drawings are given in Figures 5 (e)-(f). Patch ``N1" is
the main sunspot with negative polarity and patch ``P1" the remote
faculae with positive polarity (see Figures 1 and 4). They are
connected by large loops, which can be identified in the AIA EUV
images (see Figures 3 and 4). Patches ``N2" (``N3") and ``P2"
(``P3") are the opposite polarities of dipole ``1" (``2") that
emerged near the negative sunspot, and are connected by the initial
loops emerging from the sub-photosphere (see Figure 2). There also
exists a connection between patch ``P2" and the nearby ``N4" (also
see loop ``L3" in Figure 2). After the emergence of dipole ``1", the
negative patch ``N2" moved toward the negative sunspot ``N1" (panel
(e)), and then merged with it (panel (f)). During this process,
patch ``N2" performed a clockwise rotation around ``N1" (see Movie
2), which in our opinion led to the reconnection between a set of
small loops connecting ``N2" and ``P2" and a set of large ones
connecting ``N1" and ``P1". When the ejected material reached the
remote positive faculae ``P1", chromospheric brightenings can be
observed. In fact, these two sets of loops were distributed in
patches ``N1" (actually containing a small patch ``N2"), ``P1", and
``P2", forming a ``three-legged" structure. According to previous
studies, magnetic reconnection in this kind of configuration is much
efficient in leading to a confined flare (e.g., Hanaoka 1996, 1997).
Magnetic reconnection also occurred between two groups of small
loops, one group connecting patches ``N3" and ``P3" and the other
group connecting patches ``N4" and ``P2". Due to the reconnection,
new connections (``N3"--``P2", ``N4"--``P3") were formed (also see
Figure 2).

As revealed in previous observations, the overlying loops play an
important role in resulting in failed filament eruptions (e.g., Ji
et al. 2003; Chen et al. 2013). In this study, many pre-existing
large loops above the confined flares are clearly observed in AIA
EUV lines, e.g., 171 {\AA}, and 131 {\AA} (see Figures 3 and 4).
Although the large loops could not be identified in the H$\alpha$ at
the pre-flare stage, they were tracked by dark materials during the
flares. The dark materials were ejected toward the remote positive
faculae, appearing as dark arcades (see Figures 4 and 5). In our
opinion, the overlying large-scale loops contribute to the
confinement of flares, due to the existence of magnetic tension of
the overlying loops. In addition, since the three flares in this
study are C-class ones, the flare energy is not high enough to
destroy the overlying loops greatly. The three homologous confined
flares occurred successively, implying that, during each flare
process, only a part of loops were involved in the reconnection and
the main structure was not destroyed. As new magnetic flux
continuously emerged, another instability was produced, thus
resulting in a new reconnection with energy release (Zhang \& Wang
2002).

\acknowledgments { We thank the referee for constructive comments
and valuable suggestions. The data are used courtesy of NVST and SDO
science teams. This work is supported by the Outstanding Young
Scientist Project 11025315, the National Basic Research Program of
China under grant 2011CB811403, the CAS Project KJCX2-EW-T07, the
National Natural Science Foundations of China (11203037, 11221063,
11373004, and 11303049), and the Strategic Priority Research
Program--The Emergence of Cosmological Structures of the Chinese
Academy of Sciences (No. XDB09000000).}

{}

\end{document}